\documentclass[aps,prl,showpacs,groupedaddress,floatfix,nofootinbib]{revtex4}
\usepackage{amsmath}

\usepackage{graphicx}

\usepackage{color}

\def\beq{\begin{eqnarray}}
\def\eeq{\end{eqnarray}}

\begin{document}

\title{Systematic perturbation calculation of integrals with applications to physics}

\author{Paolo Amore}\email{paolo@cgic.ucol.mx}

\author{Alfredo Aranda}%\email{fefo@ucol.mx}

\affiliation{Facultad de Ciencias, Universidad de Colima, \\
Bernal D\'{\i}az del Castillo 340, Colima, Colima, Mexico}

\author{Francisco M. Fern\'andez}\email{fernande@quimica.unlp.edu.ar}

\affiliation{INIFTA (Conicet,UNLP), Diag. 113 y 64 S/N, \\
Sucursal 4, Casilla de Correo 16, 1900 La Plata, Argentina}

\author{Ricardo S\'aenz}%\email{rasaenz@ucol.mx}
\affiliation{Facultad de Ciencias, Universidad de Colima, \\
Bernal D\'{\i}az del Castillo 340, Colima, Colima, Mexico}

\begin{abstract}
In this paper we generalize and improve a method for calculating
the period of a classical oscillator and other integrals of
physical interest, which was recently developed by some of the
authors. We derive analytical expressions that prove to be more
accurate than those commonly found in the literature, and test the
convergence of the series produced by the approach.
\end{abstract}

\maketitle

\section{Introduction}

\label{sec:intro}

There is great interest in the development of new methods for the treatment
of nonlinear problems. For example, Amore and S\'{a}enz~\cite{AS:04} have
recently considered the problem of calculating the period of a classical
oscillator with high precision. The method that they propose proves to be
quite effective and applicable with limited effort to a large spectrum of
problems. In fact, some of the problems that considered in ref. \cite{AS:04}
are textbook examples, for which such method provides very accurate
solutions; remarkably such solutions do not involve complicated
transcendental functions but are expressed in terms of elementary functions.
Amore and collaborators have recently discussed mainly two strategies for
solving nonlinear problems. One of them is the direct treatment of the
integral that gives the period of oscillation or any other relevant property
of the system, and the other is based on an improved Lindstedt--Poincar\'{e}
technique \cite{AA1:03, AA2:03, AM:04, AS:04}. In both cases the authors
resort to a sort of variational perturbation theory like that often used in
quantum mechanics and other fields of theoretical physics to treat divergent
series \cite{AFC90}.

The purpose of the present paper is twofold: in first place we want to
generalize the method of \cite{AS:04} and to express it in a more systematic
fashion; in second place we want to extend the previous analysis to consider
large orders and discuss the convergence of the expansions that we obtain.
We investigate the systematic calculation of integrals with applications in
various fields of theoretical physics, such as, for example, the period of
nonlinear oscillations, the deflection of the light by the sun, and the
precession of the perihelion of a planet orbiting around the sun. We try to
provide simple though sufficiently accurate analytical formulas, and test
the convergence of the series, which give such formulas at low order. In
doing so, we compare our results with those in recent literature.

\section{The method}

\label{sec:method}

Many physical problems reduce to the calculation of integrals of the kind:
\begin{equation}
I=\int_{x_{-}}^{x_{+}}\ \frac{dx}{\sqrt{Q(x)}}\ ,
\end{equation}
where $Q(x)$ has simple zeros at $x_{-}$ and $x_{+}$ and is nonnegative in
the interval $x_{-}<x<x_{+}$. Such integrals appear in many branches of
classical mechanics as we will shortly show. In order to derive a simple
analytical expression for the integral $I$ we add and subtract a function $%
Q_{0}(x)$, which satisfies the same boundary conditions, and write
\begin{equation}
I=\int_{x_{-}}^{x_{+}}\ \frac{dx}{\sqrt{Q_{0}(x)}}\ \sqrt{\frac{1}{1+\Delta
(x)}}\ ,  \label{eq:I_Delta}
\end{equation}
where $\Delta (x)\equiv \frac{Q(x)-Q_{0}(x)}{Q_{0}(x)}$.

The method that we propose consists of expanding the integrand in powers of $%
\Delta (x)$ which leads to a series of the form
\begin{equation}
I=\sum_{n=0}^{\infty }I_{n}\ ,  \label{eq:I_series}
\end{equation}
where
\begin{equation}
I_{n}=\left(
\begin{array}{c}
-1/2 \\
n
\end{array}
\right) \ \int_{x_{-}}^{x_{+}}\ \frac{\Delta ^{n}(x)}{\sqrt{Q_{0}(x)}}\ dx\
\label{eq:I_n}
\end{equation}
and $\left(
\begin{array}{c}
a \\
b
\end{array}
\right) =a!/[b!(b-a)!]$ is a combinatorial number. Present method proves to
be practical when we can obtain simple analytical solutions for a
sufficiently large number of integrals in (\ref{eq:I_n}).

\section{Harmonic approximation}

\label{sec:HA}

According to what was said above about the function $Q(x)$ we can write
\begin{equation}
Q(x)=R(x)\ (x-x_{-})\ (x_{+}-x)\ ,
\end{equation}
where $R(x)\geq 0$ in $x_{-}<x<x_{+}$. A simple suitable reference function $%
Q_{0}(x)$ for many physical problems is
\begin{equation}
Q_{0}(x)=\omega ^{2}\ (x-x_{-})\ (x_{+}-x)
\end{equation}
where $\omega $ is an adjustable parameter, so that
\begin{equation}
\Delta (x)=\frac{R(x)-\omega ^{2}}{\omega ^{2}}\ .
\end{equation}

In order to simplify the equations still further, we introduce the change of
variable
\begin{equation}
x=\frac{x_{+}+x_{-}}{2}+\frac{x_{+}-x_{-}}{2}\ \cos \theta
\label{eq:x(theta)}
\end{equation}
where $0\leq \theta \leq \pi $. We can thus rewrite the integral (\ref
{eq:I_Delta}) as
\begin{equation}
I=\frac{1}{\omega }\ \int_{0}^{\pi }\ \frac{d\theta }{\sqrt{1+\Delta }}
\label{eq:I_theta}
\end{equation}
and the terms in the expansion (\ref{eq:I_n}) become
\begin{equation}
I_{n}=\frac{1}{\omega }\ \left(
\begin{array}{c}
-1/2 \\
n
\end{array}
\right) \ \int_{0}^{\pi }\ \Delta ^{n}\ d\theta \ .  \label{eq:I_n_theta}
\end{equation}

Clearly, the integral $I$ is independent of $\omega $, but the partial sums
\begin{equation}
S_{N}=\sum_{n=0}^{N}I_{n}  \label{eq:S_N}
\end{equation}
will depend on that arbitrary parameter. It is therefore reasonable to
require the partial sums to be independent of $\omega $, which leads to the
principle of minimal sensitivity (PMS)\cite{Ste81} that states that the
optimal value of $\omega $ should satisfy
\begin{equation}
\frac{\partial S_{N}}{\partial \omega }=0
\end{equation}
If we take into account the properties of the combinatorial numbers and that
$\frac{\partial \Delta }{\partial \omega }=-\frac{2}{\omega }\ (1+\Delta )$,
we can easily prove that
\begin{equation}
\frac{\partial S_{N}}{\partial \omega }=-\frac{2N+1}{\omega }\ I_{N}\ .
\end{equation}
According to this equation, the PMS condition is equivalent to $I_{N}=0$
which only takes place for odd values of $N$ in the real field. For even
values of $N$ we may instead resort to the alternative PMS condition $%
\partial ^{2}S_{N}/\partial \omega ^{2}=0$ but it is not necessary as we
will see later on.

\section{Anharmonic oscillators}

\label{subsec:AO}The periods of many anharmonic oscillators have been widely
studied and therefore they are suitable benchmarks for new approaches. Here
we consider a particle of unit mass moving in a one dimensional anharmonic
potential $V(x)$ and calculate the period according to the well known
expression
\begin{equation}
T=\int_{x_{-}}^{x_{+}}\frac{\sqrt{2}\ dx}{\sqrt{E-V(x)}}\ .
\end{equation}

\subsection{Duffing oscillator}

A widely studied example is the Duffing oscillator, which corresponds to the
potential $V(x)=\frac{1}{2}x^{2}+\frac{\mu }{4}\ x^{4}$. In this case we
choose $Q(x)=E-V(x)=(A^{2}-x^{2})\ \left[ \frac{1}{2}+\frac{\mu }{4}\
(A^{2}+x^{2})\right] $, where $A>0$ is the amplitude of the oscillations,
and $Q_{0}=\omega ^{2}\ (A^{2}-x^{2})$. Notice that $T=\sqrt{2}I$ and that $%
\Delta $ takes a simple form:
\begin{equation}
\Delta =\frac{1}{\omega ^{2}}\ \left[ \frac{\mu }{4}\ (A^{2}+x^{2})-\omega
^{2}+\frac{1}{2}\right] =\frac{1}{\omega ^{2}}\ \left[ \frac{\mu A^{2}}{4}\
(1+\cos ^{2}\theta )-\omega ^{2}+\frac{1}{2}\right] \ ,
\label{eq:Delta_Duffing}
\end{equation}
where $x=A\cos \theta $ as follows from equation (\ref{eq:x(theta)}) with $%
x_{+}=-x_{-}=A$. Notice that the period depends only on $\rho =\mu A^{2}$.

The value of $\omega $ according to the PMS to first order
\begin{equation}
\omega _{PMS}=\sqrt{\frac{4+3\rho }{8}}  \label{eq:omega_PMS_Duffing}
\end{equation}
yields
\begin{equation}
\Delta _{PMS}=\frac{\rho }{4+3\rho }\ \cos 2\theta \ .  \label{eq:Delta_PMS}
\end{equation}
Eq.~(\ref{eq:Delta_PMS}) shows that $|\Delta _{PMS}|<1$ for all values of $%
\theta $, and $\rho $ so that the series (\ref{eq:I_series}) converges for
all values of $\rho $.

A most interesting feature of the PMS for this model is that $I_{2n+1}\left(
\omega =\omega _{PMS}\right) =0$ for all $n=0,1,\ldots $, and for the same
value of $\omega _{PMS}$ given by eq.~(\ref{eq:omega_PMS_Duffing}). The
calculation of the even terms is straightforward and yields the compact
expression
\begin{equation}
T=\frac{4\pi }{\sqrt{4+3\rho }}\ \sum_{n=0}^{\infty }\ (-1)^{n}\ \left(
\begin{array}{c}
-1/2 \\
n
\end{array}
\right) \ \left(
\begin{array}{c}
-1/2 \\
2n
\end{array}
\right) \xi ^{2n},\;\xi =\ \frac{\rho }{4+3\rho }\ .
\label{eq:T_series_Duffing_present}
\end{equation}

If we choose $\omega ^{2}=1+\rho $ in eq.~(\ref{eq:Delta_Duffing}) then eq.~(%
\ref{eq:I_theta}) gives us a well--known exact expression for the period\cite
{Nayfeh81}
\begin{equation}
T_{exact}=\frac{4}{\sqrt{1+\rho }}\ \int_{0}^{\pi /2} \frac{d\alpha }{\sqrt{%
1-\kappa \ \sin ^{2}\alpha }},  \label{eq:Texact_Duffing}
\end{equation}
where $\kappa =\frac{\rho }{2(1+\rho )}$. We thus obtain the alternative
series expansion:
\begin{equation}
T=\frac{2\pi }{\sqrt{1+\rho }}\ \sum_{n=0}^{\infty }\left(
\begin{array}{c}
-1/2 \\
n
\end{array}
\right) ^{2}\ \kappa ^{n}  \label{eq:T_series_Nayfeh}
\end{equation}

If $\mu <0$, then the potential of the oscillator exhibits two barriers and
the amplitude of the motion cannot be larger than $A_{L}=1/\sqrt{-\mu }$.
Consequently, the exact expression for the period is valid for $\rho >-1$.
Present series (\ref{eq:T_series_Duffing_present}) converges uniformly for $%
\rho >-1$ whereas the series in eq.~(\ref{eq:T_series_Nayfeh}) does not
converge for $-1<\rho <-2/3$.

The exact solution $x(t)$ for the Duffing oscillator satisfies the virial
theorem $\overline{\dot{x}^{2}}=\overline{x^{2}}+\mu \overline{x^{4}}$,
where $\overline{f(x)}$ stands for the classical expectation value of $%
f(x(t))$. It is most interesting to note that the value of $\omega $ in $%
x_{0}(t)=A\cos (\omega t+\phi )$ that makes $x_{0}(t)$ to satisfy the virial
theorem for the Duffing oscillator is $\sqrt{2}\omega _{PMS}$.

In a recent paper Pelster et al. \cite{Pel:03} (PKS) calculated the leading
term of the strong--coupling expansion for the frequency of the Duffing
oscillator by means of a series produced by the Lindstedt--Poincar\'{e}
method with an adjustable harmonic frequency. In the limit $\mu \rightarrow
\infty $ they expand the frequency for unit amplitude as
\begin{equation}
\omega =\sqrt{\mu }\ \left[ b_{0}+\frac{b_{1}}{\mu }+\frac{b_{2}}{\mu ^{2}}
+\dots \right] \ .
\end{equation}

\begin{figure}[tbp]
\begin{center}
\includegraphics[width=8cm]{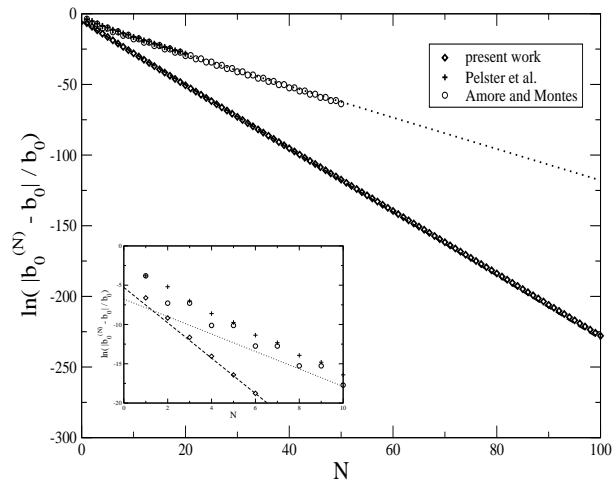}
\end{center}
\caption{Logarithmic plot of the error $|b_0^{(N)}-b_0|/b_0$ as a function
of the order $N$. Diamonds and crosses correspond to present results and
those of Pelster et al. respectively. The dashed and dotted lines correspond
to the linear fits of each data. The circles correspond to the results of
\protect\cite{AM:04}.}
\label{Fig_1}
\end{figure}

If $T^{(N)}$ is the Nth--partial sum for the series (\ref
{eq:T_series_Duffing_present}) we obtain the coefficient $b_{0}^{(N)}$ to
order $N$ as:
\begin{equation}
b_{0}^{(N)}=\lim_{\mu \rightarrow \infty }\frac{2\pi }{\sqrt{\mu }T^{(N)}}=
\frac{\sqrt{3}}{2\sum_{j=0}^{N}\left( \frac{-1}{9}\right) ^{j}\left(
\begin{array}{c}
-1/2 \\
j
\end{array}
\right) \ \left(
\begin{array}{c}
-1/2 \\
2j
\end{array}
\right) }.  \label{eq:b0_present}
\end{equation}
Fig.~\ref{Fig_1} shows the logarithmic relative error calculated with this
expression, with the values of $b_{0}^{(N)}$ given by Pelster et al. \cite
{Pel:03} and the results of Amore et al \cite{AA1:03,AM:04}. We also show
the linear fits for the first two sets of data.

Our series has by far the best rate of convergence. In fact, altough all
three series exhibit exponential convergence
\begin{equation}
\left| \frac{b_{0}^{(N)}-b_{0}}{b_{0}}\right| =e^{-\alpha -\beta N}
\end{equation}
the slope $\beta $ of our linear fit is much greater: $\beta ^{present}=\ln
(9)\approx 2.1972$, compared to $\beta ^{PKS}\approx 1.11$ \cite{Pel:03}.
Moreover, our expression (\ref{eq:b0_present}), which is much simpler than
the one of Pelster et al \cite{Pel:03}, enables us to calculate the slope $%
\beta $ exactly.

\begin{figure}[tbp]
\begin{center}
\includegraphics[width=8cm]{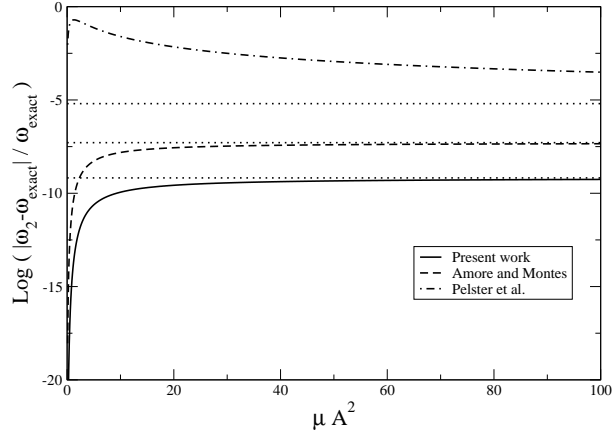}
\end{center}
\caption{Logarithmic plot of the error $|\omega_2-\omega_{exact}|/
\omega_{exact}$ as a function of the order $\mu A^2$. The three curves
correspond to the second order results for the present method, the LPLDE
method of \protect\cite{AA1:03,AM:04} and the method of Pelster et al.
respectively. The horizontal lines are the asymptotic values of the errors
(see Fig. \ref{Fig_1}).}
\label{Fig_2}
\end{figure}

In Fig.~\ref{Fig_2} we compare the logarithmic relative error for the
frequency calculated to second order with the three methods as a function of
$\rho $. The results for the method of Pelster et al. are obtained by means
of eq.~(42) of their paper \cite{Pel:03}. In the limit $\rho \rightarrow
\infty $ one recovers the asymptotic error of Fig.~\ref{Fig_1},
corresponding to $N=2$. It is interesting to notice that both the present
expansion and the result of Amore et al \cite{AA1:03,AM:04} yield an error
which is always smaller than the asymptotic one; on the other hand, the
results of Pelster et al \cite{Pel:03} do not follow this rule and yield a
particularly large error in the region of $\rho \approx 1$. We clearly
appreciate that our proposal is much more convenient than Kleinert's
square--root trick \cite{Pel:03}.

\subsection{Quadratic-sextic oscillator}

We also consider the potential $V(x)=\frac{1}{2}x^{2}+\frac{\mu }{6}x^{6}$
because it has recently been treated by means of the combination of the
methods of linearization and harmonic balance by Wu and Li \cite{WL:01}
(WL). These authors carry out calculations of low order; their best approach
is
\begin{equation}
T^{WL}=\frac{24\pi }{\sqrt{80+50\rho +\sqrt{4096+5120\rho +925\rho ^{2}}}}
,\; \rho =\mu A^{4}  \label{eq:T_WL}
\end{equation}

If we choose $\omega =\sqrt{\frac{\rho +3}{6}}$ in eq.~(\ref{eq:I_theta}) we
obtain an exact expression for the period:
\begin{equation}
T^{exact}=\frac{2\sqrt{3}}{\sqrt{\rho +3}}\int_{0}^{\pi }\frac{d\theta }{%
\sqrt{1+\frac{\rho }{\rho +3}\left( \cos ^{2}\theta +\cos ^{4}\theta \right)
}}
\end{equation}
In this case the greatest amplitude for $\mu <0$ satisfies $\rho _{L}=\mu
A_{L}^{4}=-1$.

The application of our method is similar to that for the Duffing oscillator
discussed above. We obtain:
\begin{equation}
\Delta =\frac{3+\rho -6\,\omega ^{2}+\rho \,\cos (\theta )^{2}+\rho \,\cos
(\theta )^{4}}{6\,{\omega }^{2}}
\end{equation}
In order to keep our equations as simple as possible we use the optimal
value of $\omega $ of first order to all orders:
\begin{equation}
\omega _{PMS}=\frac{\sqrt{5\rho +8}}{4}
\end{equation}

which leads to
\begin{equation}
\Delta _{PMS}=\frac{\rho }{3(5\rho +8)}\ \left[ 8\ \cos 2\theta +\cos
4\theta \right] \ .
\end{equation}

The expressions thus obtained by our method through order three are not as
accurate as the one of Wu and Li (\ref{eq:T_WL}). At fourth order we have a
formula that is not more complicated than eq.~(\ref{eq:T_WL}) and is
certainly more accurate for all values of $\rho $:
\begin{equation}
T^{(4)}=\frac{\sqrt{2}\pi \left( 6097185\rho ^{4}+37821440\rho
^{3}+89272320\rho ^{2}+94371840\rho +3774873\right) }{2304\left( 5\rho
+8\right) ^{9/2}}.
\end{equation}
Notice, for example that
\[
\lim_{\rho \rightarrow \infty }\sqrt{\rho }T^{exact}=8.413092631\ \ \ ,\ \ \
\lim_{\rho \rightarrow \infty }\sqrt{\rho }T^{WL}=8.4081\ \ \ ,\ \ \
\lim_{\rho \rightarrow \infty }\sqrt{\rho }T^{(4)}=8.41292
\]
and that
\[
T^{exact}(\rho =-0.9)=10.93467798\ \ \ ,\ \ \ T^{WL}(\rho =-0.9)=10.62\ \ \
,\ \ \ T^{(4)}(\rho =-0.9)=10.67\ .
\]

Our method enables us to derive explicit equations at all orders. If we
calculate the integrals in eq.~(\ref{eq:I_n_theta}) exactly, we obtain
\[
T=\sum_{n=0}^{\infty }\frac{4\pi }{\sqrt{5\rho +8}}\ \left(
\begin{array}{c}
-1/2 \\
n
\end{array}
\right) \ \left[ \frac{\rho }{3(5\rho +8)}\right] ^{n}\ \mathcal{J}_{n}
\]
where
\[
\mathcal{J}_{n}\equiv \sum_{k_{1}=0}^{n}\ \sum_{k_{2}=0}^{k_{1}}\
\sum_{k_{3}=0}^{n-k_{1}}\ 2^{3k_{1}-n}\ \left(
\begin{array}{c}
n \\
k_{1}
\end{array}
\right) \ \left(
\begin{array}{c}
k_{1} \\
k_{2}
\end{array}
\right) \ \left(
\begin{array}{c}
n-k_{1} \\
k_{3}
\end{array}
\right) \ \delta _{k1,2n-2k_{2}-4k_{3}}\ .
\]

\begin{figure}[tbp]
\begin{center}
\includegraphics[width=8cm]{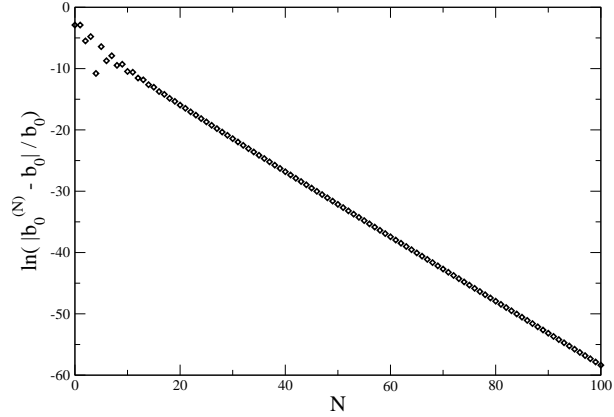}
\end{center}
\caption{Logarithmic plot of the error $|b_0^{(N)}-b_0|/b_0$ as a function
of the order $N$ for the quadratic sextic oscillator.}
\label{Fig_3}
\end{figure}

In Fig.~(\ref{Fig_3}) we plot the logarithmic relative error of the period
for $\rho \rightarrow \infty $ as a function of the order of approximation.
Once again we see that our expansion converges exponentially.
Indeed, the error decreases as $e^{-\alpha-\beta N}$, where it is not hard to see 
that $\beta = \ln (5/3) \approx .5108$.

\subsection{Other parity--invariant anharmonic oscillators}

The anharmonic oscillators with even potentials of the form
\begin{equation}
V(x)=\frac{x^{2}}{2}+\frac{\mu x^{2K}}{2K},\;K=2,3,\ldots  \label{eq:Vx2x2K}
\end{equation}
exhibit many features in common. For example, the period $T$ depends on the
parameter $\rho =\mu A^{2K-2}$ and satisfies a small--$\rho $ expansion
\begin{equation}
T=\sum_{j=0}^{\infty }a_{j}\rho ^{j}
\end{equation}
and a large--$\rho $ expansion
\begin{equation}
T=\frac{1}{\sqrt{\rho }}\sum_{j=0}^{\infty }c_{j}\rho ^{j}.
\end{equation}
The optimal value of $\omega $ given by our method is of the form
\begin{equation}
\omega ^{2}=\frac{1+\kappa \rho }{2}.
\end{equation}
The optimal value of $\kappa $ is the same to all orders for $K=2$ as we
have already discussed above. It changes slowly with $N$ for the sextic
oscillator and more rapidly for greater values of $K$. In order to make the
discussion simpler we just consider the convergence of the series for the
most difficult case given by
\begin{equation}
c_{0}=\lim_{\rho \rightarrow \infty }\sqrt{\rho }T
\end{equation}
For $K=5$ the value of $\kappa $ given by the PMS of first order does not
give us a convergent series for this quantity. The reason is that the
maximum value of $|\Delta |$ is greater than 1 under such condition. Values
of $\kappa $ given by the PMS of higher order correct this problem.
Numerical calculation suggests that the values of $\kappa $ given by the PMS
at increasingly greater order approaches the value of $\kappa $ for which
the maximum of $\Delta $ equals minus one times its minimum. Such value of $%
\kappa $ makes $|\Delta |<1$ for all values of $\theta $ and we thus have a
remarkably simple criterion for obtaining a convergent series. For
concreteness we call $\kappa _{b}$ this value of $\kappa $ because it
balances the positive and negative values of $\Delta $ in such a way that $%
|\Delta |<1$. We find that $\kappa _{b}$ is suitable for high--order
calculations but it is not necessarily the most convenient one for the
derivation of simple and accurate analytical expressions like those obtained
from the PMS of low order.

Plots of $c_{0}^{(2N+1)}$ vs. $\kappa $ obtained by partial sums of order $%
2N+1$ show that the PMS always give the best approach to the exact value at
each order. Plots of $\log \left| \frac{c_{0}^{(2N+1)}-c_{0}}{c_{0}}\right| $
vs. $N$ suggest that our partial sums with $\kappa _{b}$ converge
exponentially for all anharmonic oscillators (\ref{eq:Vx2x2K}).

For negative values of $\mu $ there is periodic motion provided that $\rho
>-1$. Plots of $\log \left| \frac{T^{(N)}-T}{T}\right| (\rho =-0.9)$ vs $N$
for anharmonic oscillators with $K=3,4,5$ show two almost straight lines,
one for even $N$ lying always lower than the one for odd $N$. In other
words, the partial sums with even $N$ and odd $N$ converge exponentially to
the actual value of $T(\rho =-0.9)$.

\subsection{Quadratic cubic oscillator}

The treatment of noneven potentials is slightly different. We illustrate the
procedure by means of the simplest such oscillator:
\begin{equation}
V(x)=\frac{x^{2}}{2}+\mu \frac{x^{3}}{3}.
\end{equation}

The factorization of $Q(x)=E-V(x)$ leads to
\begin{eqnarray}
Q(x) &=&(x-x_{-})(x_{+}-x)(b_{0}+b_{1}x) \ ,
\end{eqnarray}
where
\begin{eqnarray}
b_{0} = -\frac{x_{+}x_{-}}{2(x_{+}^{2}+x_{+}x_{-}+x_{-}^{2})} \ \ \ , \ \ \
b_{1} =-\frac{x_{+}+x_{-}}{2(x_{+}^{2}+x_{+}x_{-}+x_{-}^{2})}
\end{eqnarray}
and
\begin{eqnarray}
\mu = - \frac{3}{2} \ \frac{x_-+x_+}{x_+^2+x_+ x_- + x_+^2} \ .
\end{eqnarray}

A straightforward calculation based on the change of variable (\ref
{eq:x(theta)}) shows that the PMS at first order yields
\begin{equation}
\omega =\frac{\sqrt{\left( -x_{+}^{2}-4x_{+}x_{-}-x_{-}^{2}\right) }}{ 2%
\sqrt{x_{+}^{2}+x_{+}x_{-}+x_{-}^{2}}}
\end{equation}
and
\begin{equation}
\Delta =\frac{(x_{+}^{2}-x_{-}^{2})\cos \theta }{%
x_{+}^{2}+4x_{+}x_{-}+x_{-}^{2}}
\end{equation}

It is clear that all the perturbation corrections of odd order vanish and we
obtain the same series as in the case of the Duffing oscillator:
\begin{equation}
T=\frac{\sqrt{2}\pi }{\omega }\sum_{j=0}^{\infty }(-1)^{j}\left(
\begin{array}{c}
-1/2 \\
j
\end{array}
\right) \left(
\begin{array}{c}
-1/2 \\
2j
\end{array}
\right) \xi ^{2j},\;\xi =\frac{(x_{+}^{2}-x_{-}^{2})}{%
x_{+}^{2}+4x_{+}x_{-}+x_{-}^{2}}  \label{eq:Tx2x3_series}
\end{equation}

The potential--energy function exhibits a maximum $V(x_{M})=1/(6\mu ^{2})$
at $x_{M}=-1/\mu $;\ therefore, there will be periodic motion for all values
of energy satisfying
\begin{equation}
E<\frac{1}{6\mu ^{2}}=\frac{2\left( x_{+}^{2}+x_{+}x_{-}+x_{-}^{2}\right)
^{2}}{27\left( x_{+}+x_{-}\right) ^{2}}.
\end{equation}

In this case we obtain the exact solution by means of the integral
\begin{equation}
T=\frac{\sqrt{2}}{\omega }\int_{0}^{\pi }\frac{d\theta }{\sqrt{1+\xi \cos
\theta }}
\end{equation}
that yields the series (\ref{eq:Tx2x3_series}) on expansion in a Taylor
series about $\xi =0$.

An interesting application of the main equations for the quadratic--cubic
anharmonic oscillator is the precession of the perihelion of a planet. As
shown in a previous paper by some of the authors~\cite{AS:04} the expression
for the angular precession $\Delta \phi $ that appears in most textbooks can
be rewritten as
\begin{equation}
\Delta \phi = 2\int_{z_{-}}^{z_{+}}\frac{dz}{\sqrt{\left( z_{+}-z\right)
\left( z-z_{-}\right) \left[ 1-2GM\left( z+z_{-}+z_{+}\right) \right] }}%
-2\pi,  \label{eq:ang_prec}
\end{equation}
where, $z=1/r$, $z_{\pm }=1/r_{\pm }$ and $r_{-}<r_{+}$ are the shortest
(perihelia) and largest (aphelia) distances from the sun. Taking into
account the expressions for the quartic--cubic oscillator derived above we
have
\begin{equation}
\omega =\sqrt{1-3GM\left( z_{+}+z_{-}\right) }=\sqrt{1-\frac{6GM}{L}},
\label{eq:omega}
\end{equation}
where the semilatus rectum $L$ is given by $1/L=\left( z_{+}+z_{-}\right) /2$%
. We also obtain
\begin{equation}
\Delta =\xi \cos \theta ,\;\xi =\frac{GM\left( z_{+}-z_{-}\right) }{%
3GM\left( z_{+}+z_{-}\right) -1}=\frac{GM\sqrt{\frac{a-L}{a}}}{6GM-L},
\end{equation}
where $a=\left( 1/z_{+}+1/z_{-}\right) /2$ is the semimajor axis of the
ellipse. We thus obtain the series
\begin{equation}
\Delta \phi =2\pi \left[\frac{1}{\omega }\sum_{j=0}^{\infty }(-1)^{j}\left(
\begin{array}{c}
-1/2 \\
j
\end{array}
\right) \left(
\begin{array}{c}
-1/2 \\
2j
\end{array}
\right) \xi ^{2j} -1 \right]  \label{eq:precession}
\end{equation}
that is a generalization to all orders of an expression of second order
developed by Amore and S\'{a}enz \cite{AS:04}.

\begin{figure}[tbp]
\begin{center}
\includegraphics[width=8cm]{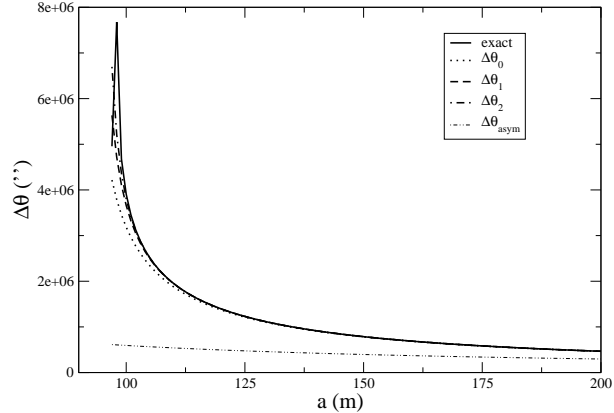}
\end{center}
\caption{Precession of the orbit of a planet (in seconds of degree)
calculated assuming $M = 1.97 \times 10^{30} \ kg$, $G/c^2 = 7.425 \times
10^{-30} \ m/kg$ and $\epsilon = 0.2506$ as a function of $a$ (i.e. the
average of perihelion and aphelion) for values close to a black hole. The
solid line is the exact result; the thin line corresponds to the asymptotic
formula; the other lines correspond to the different orders to which our
approximation has been applied.}
\label{Fig_4}
\end{figure}

\begin{figure}[tbp]
\begin{center}
\includegraphics[width=8cm]{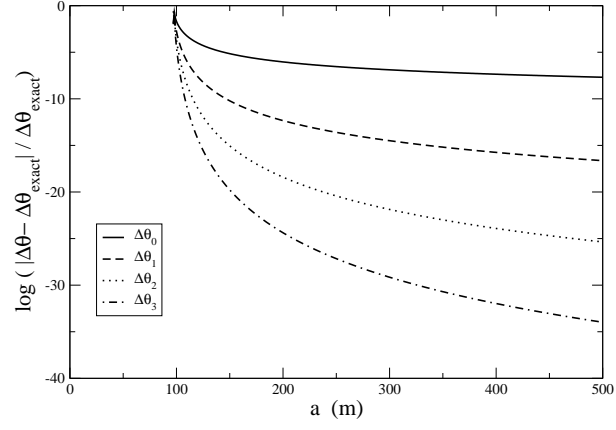}
\end{center}
\caption{Logarithmic plot of the error over the precession of the perihelion
as a function $a$. The different lines correspond to the different orders to
which our approximation has been applied.}
\label{Fig_5}
\end{figure}

In Fig.~\ref{Fig_4} we plot the angular precession of the orbit of a planet
obtained using the exact formula and the approximations given by partial
sums of our series (\ref{eq:precession}) for values of $a$ very close to the
Schwartzchild radius. In Fig.~\ref{Fig_5} we plot the logarithm of the
relative error over the angular precession. We notice that our approximation
is able to reproduce the exact result very accurately, even for very small $a
$. However, our series does not reproduce the singular point of the exact
integral exactly, because this singularity is due to the appearance of the
third root of the cubic polynomial inside the integration interval. In other
words, the factor function $R(\theta )$ is negative for some values of $%
\theta $ when $a$ is smaller than the critical value $a_{c}=97.9173$. In
spite of this limitation, the partial sums of our series approach the exact
value of the integral as we add more terms. We can improve our results by
splitting the integral into two parts, exactly at the new zero of $Q(z)$,
treating each integral in the way indicated. However, we do not deem it
necessary to discuss such calculation here because it does not add anything
new to what was already done.

We can apply present method to anharmonic oscillators with more than two
terms in the potential--energy function; as an example consider
\[
V(x)=a_{2}x^{2}+a_{3}x^{3}+a_{4}x^{4}.
\]
If we write
\[
Q(x)=E-V(x)=\left( x-x_{-}\right) \left( x_{+}-x\right) \left(
b_{0}+b_{1}x+b_{2}x^{2}\right)
\]
then we obtain
\begin{eqnarray*}
E &=&-x_{+}x_{-}\left[ a_{2}+a_{3}\left( x_{+}+x_{-}\right) +a_{4}\left(
x_{+}^{2}+x_{+}x_{-}+x_{-}^{2}\right) \right] \\
b_{0} &=&a_{2}+a_{3}\left( x_{+}+x_{-}\right) +a_{4}\left(
x_{+}^{2}+x_{+}x_{-}+x_{-}^{2}\right) \\
b_{1} &=&a_{3}+a_{4}\left( x_{+}+x_{-}\right) \\
b_{2} &=&a_{4}
\end{eqnarray*}
and
\[
\Delta =\frac{b_{0}+b_{1}x+b_{2}x^{2}-\omega ^{2}}{\omega ^{2}}.
\]

It follows from the PMS condition $I_{1}=0$ that
\[
\omega _{PMS}=\frac{\sqrt{16b_{0}+8b_{1}\left( x_{-}+x_{+}\right)
+2b_{2}\left( 3x_{-}^{2}+2x_{+}x_{-}+3x_{+}^{2}\right) }}{4}.
\]
The expression of the period to order zero is quite simple $T\approx \sqrt{2}
I_{0}=\sqrt{2}\pi /\omega $ but the expression to second order $T\approx
\sqrt{2}\left( I_{0}+I_{2}\right) $, although simple, is rather long to be
shown here.

\subsection{The simple pendulum}

The method just outlined also applies to non polynomial potentials. As a
representative example we consider the simple pendulum, and without loss of
generality we choose unit mass and length. We obtain particularly simple
expressions if we expand the potential
\[
V(\phi )=1-\cos \phi
\]
in a Taylor series. For example, in the first approximation $V(\phi )\approx
\phi ^{2}/2$ is harmonic and the period is independent of the amplitude $A$.
The resulting textbook expression for the period is valid for small values
of $A$. In order to test the accuracy of the approximations we consider the
exact expression for the period
\[
T=2\int_{0}^{\pi }\frac{d\alpha }{\sqrt{1-\sin ^{2}\frac{A}{2}\,\sin ^{2}
\frac{\alpha }{2}}}.
\]

If we keep one more term in the expansion $V(\phi )\approx \phi ^{2}/2-\phi
^{4}/24$ we can apply all the results for the Duffing oscillator with $\mu
=-1/6$. The leading term of the series (\ref{eq:T_series_Duffing_present})
yields in this case
\[
T=\frac{4\sqrt{2}\pi }{\sqrt{8-A^{2}}}.
\]
Taking into account terms of higher order of the series (\ref
{eq:T_series_Duffing_present}) does not improve results because the mayor
source of error is the expansion of the potential only through fourth order.

The next approximation in the expansion of the potential is $V(\phi )\approx
\phi ^{2}/2-\phi ^{4}/24+\phi ^{6}/720$. Present method yields
\begin{equation}
\omega =\frac{\sqrt{6}\left( A^{4}-24A^{2}+192\right) }{48}
\end{equation}
at first order of PMS. Keeping just the leading term of the series (\ref
{eq:I_series}) we obtain
\begin{equation}
T=\frac{16\sqrt{3}\pi }{\sqrt{A^{4}-24A^{2}+192}}.
\end{equation}
This expression gives better results than the preceding one, and in this
case the term of second order improves the accuracy (the term of first order
is zero because of the PMS). We thus obtain
\[
T=\frac{\sqrt{3}\pi \left(
253A^{8}-11904A^{6}+233280A^{4}-2211840A^{2}+8847360\right) }{15\left(
A^{4}-24A^{2}+192\right) }.
\]

\section{Conclusions}

In this paper we propose a technique for solving a class of integrals with
applications in theoretical physics. We provide a systematic general method
for writing those integrals in a form convenient for exact calculation and
for their expansion in series. We show that the resulting series are
suitable for all values of the relevant physical parameters, from unstable
equilibrium through the strong--coupling, and for all the anharmonic
oscillators considered here. We easily derive simple low--order analytical
expressions that are more accurate than those in the literature, and we have
shown that the rate of convergence of our series is greater than those
proposed by other authors. In some cases the greater convergence rate
observed is probably due to the fact that the PMS works better on the period
than on the frequency.

The approach that we propose in this paper is closely related to what is
known as variational perturbation method, which was extensively applied to
quantum mechanics and other fields of theoretical physics \cite{AFC90} (see
also additional references cited in \cite{Pel:03}). In particular, the
method of nonlinear mappings \cite{AFC90}, which focuses only on the series
itself, is more general that the standard variational perturbation theory.
The nonlinear mappings can also be applied to the problems considered here.
However, we have resorted to the technique outlined above because it has
facilitated the analysis of the convergence of the series by focusing on
properties of the integrals.

\bigskip

\begin{acknowledgments}
P.A. acknowledges support of Conacyt grant no. C01-40633/A-1.
\end{acknowledgments}

\end{document}